**Matthias Risch**    Hochschule Augsburg, Germany, University of Applied Sciences

# Can Physics Teaching be improved by Explanation of Tricks with a Motorcycle?

**Abstract**

A priority of physics instruction is to help students make the connection between the formulae they think they are required to memorize and the real world in which they interact every day. If you ask students to describe a situation "in real life" involving physical principles, the most commonly cited examples will pertain to vehicular motion [Smith 2006]. One situation "in real life" involving physical principles is vehicle dynamics. Even students who have little interest in physics eagerly discuss problems like how much a car can decelerate travelling in a flat turn or how tricks like the ´Wheely´ can be performed on a motorcycle. In the physics classroom, the motion of automotive vehicles is probably the most interesting manifestation of the principles of physics.

The laws of physics limiting movements of vehicles are deduced here in a simple derivation suited for classroom demonstration as well as for homework [Lorenzo, 2005]. Due to limits on frictional forces there are subsequent limits for acceleration, deceleration and speed in a flat turn. Frictional forces also determine the behaviour of a vehicle at rapid speed in a turn.

# 1. Maximum acceleration and maximum deceleration of vehicles

Most students have had the experience that too much deceleration of a car as well as on a bicycle will lock the wheels, prevent them from rolling and will inevitably result in a tangential motion and eventually make the vehicle leave the road. Likewise, increasing acceleration too much can make a car slide along the road leaving skid marks consisting of shredded tire surface and small melted sections of road. A car's wheels are locked when friction fails. This limit can be found by means of the free body diagram of the wheels. The static frictional force limits both acceleration as well as deceleration to maximum. The free body diagram of the wheel is considered first for acceleration. The car accelerates because a frictional force from the track acts on the rear tires to cause acceleration. For sake of simplicity only rear wheel drive is considered here, then with a few assumptions the free body diagram can be reduced to a simple level for classroom teaching without being too far from the nuts and bolts of automobile engineering; and without affecting essentials of results (figure 1)

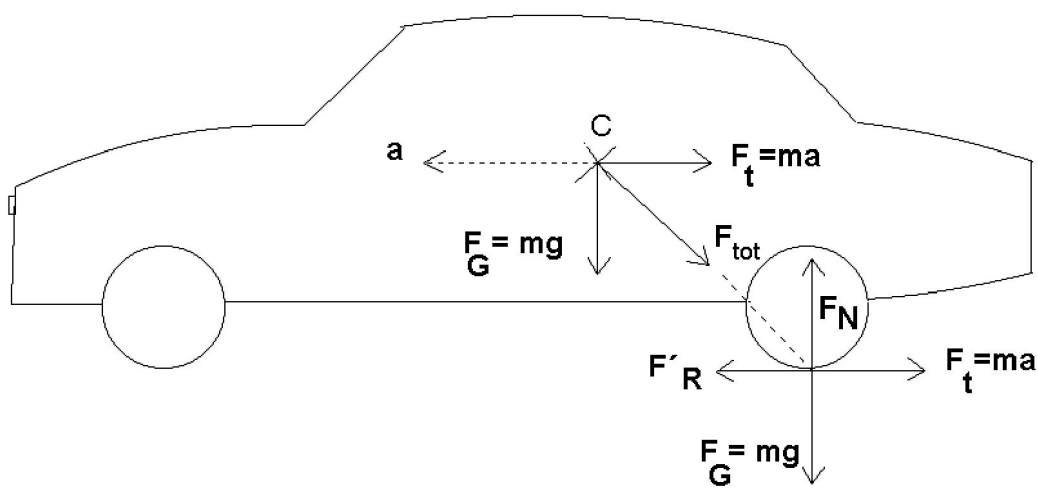

Figure 1. Simplified free body diagram of a vehicle with full acceleration, C= centre of gravity.

This simplified free body diagram neglects for example drag forces, it comprises forces in equilibrium that cancel out. Here is a description of the forces step by step:

- acceleration a is directed forward at centre of gravity C
- the accelerating force is m·a, inertial force is - m·a, both at centre of gravity C
- gravitational force is $F_G = m \cdot g$, at centre of gravity C, which forms a net force $F_{tot}$ with the inertial force -m·a
- this net force $F_{tot}$ has an angle of up to 45° to the gravitational force; in this direction of 45° lies the contact between the rear tires and the track assuming the simplifications made, such that the front tires are without any force, limiting further considerations to the rear tires
- this net force $F_{tot}$ is applied in the 45° direction to the point of contact between the rear tires and the track (dotted line)
- this net force $F_{tot}$ is split into gravitational force $F_G = m \cdot g$ and inertial force is $F_t = -m \cdot a$ acting at the point of contact between the rear tires and the track
- gravitational force $F_G = m \cdot g$ is in equilibrium with the normal force $F_N = m \cdot g$ which deforms the tire and inertial force $F_t = - m \cdot a$ is in equilibrium with static frictional force $F'_R \leq \mu' \cdot m \cdot g$ with $\mu'$ the coefficient of static friction

The static frictional force, however, cannot exceed a maximum

$$F'_{R,\,max} = \mu' \, F_N \cdot = \mu' \cdot m \cdot g \qquad \mu' = \text{coefficient of static friction}$$

For practical enduring tires this coefficient of static friction is restricted to less or equal than 1. For race cars extremely soft tires gluing to the track are used with a higher $\mu'$ wearing out after a short mileage and leaving skid marks consisting of shredded fragments of tire and small melted sections of road. This type of tire cannot be used in eve-ryday life. Therefore, the static frictional force is more or less restricted to the amount of the normal force. Since the static frictional force is in equilibrium with the accelerating force having the same amount, the accelerating force likewise is restricted to the amount of the normal force resulting in a maximum acceleration:

$$F'_{R, max} = F_{t, max} = m \cdot a_{max} = \mu' \, m \cdot g \qquad (1)$$

Since the mass m cancels out, a maximum acceleration of vehicles driven by wheels is left which is determined by coefficient of static friction µ´

$$a_{max} = \mu' \cdot g \qquad (2)$$

For good commercial tires and a dry and clean track the coefficient of static friction is 1 and maximum acceleration then is the constant of gravity g. For very high speed, this limit will be shifted because of additional forces from drag and lift caused by airflow. Relations (1) and (2) have been deduced using some assumptions concerning modern and elaborate construction of a car, such as suited position of the centre of gravity which should be in an angle of 45° above the point of contact between the rear tires and the track. Maximum acceleration is the constant of gravity g. This lacks feasibility for commercial car races because there electrical blankets, before starting, heat the tires such that they will stick at the road ripping off parts of the tire during drive which can be seen as skid-marks. Thus, after a few rounds, new tires are needed, making this technique not suited for everyday use. In "stock-car" contests the drivers will heat their tires by intentional skidding in "burn-outs". For a short period, the heated and sticky tires enable this limit of maximum acceleration (g) to be overcome, on the expense of ripped and ruined tires.

Relations (1) and (2) have a lot of practical implications for everyday driving. When the driving force and the equal amount of inertial force exceeds the limit of static frictional force, the tires will start to skid on the track and the car's wheels lock keeping them from rolling. Then the tires will transmit the kinetic frictional force rather than the static frictional force. The skidding of the tires will then produce a screeching noise and a diminished acceleration since the kinetic coefficient of friction is about 15% less than the static coefficient of friction. Since the wheels are kept from rolling the car will become unstable in its track and cannot be controlled by the steering wheel. The excess energy will heat and wear the tires excessively.

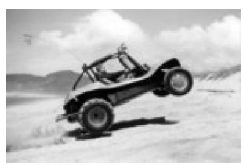

*Figure 2.    Starting of a car with maximum acceleration.*

With the maximum possible acceleration in motorcycles or bicycles, the driver can shift the centre of gravity on a position of 45° above the point of contact between the rear tires and the track as assumed in figure 1. Then the front wheel then becomes free of any part of the net force, neither inertial or gravity part of the forces will act on the front wheel as shown in figure 1. Thus, the driver can easily lift the front wheel, a trick called a ´Wheely´, used by motorcycle enthusiasts stunning spectators by using this force equilibrium and application of free body diagram (figure 3).

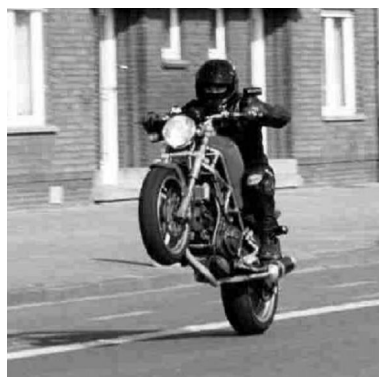

*Figure 3.    Full acceleration on a motorcycle (courtesy of a Belgian motorcycle club).*

When the vehicle is decelerating, the force of inertia is directed in the opposite direction. Again, forces of gravity and inertia are combined to give a net force, which is directed forward as seen in figure 4 [Wehrbein 2004].

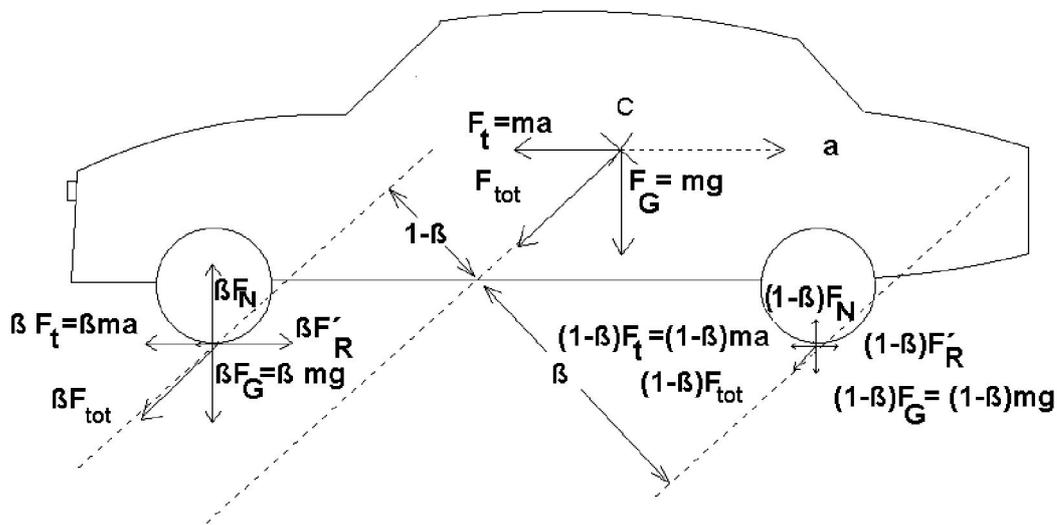

Figure 4.    Free body diagram of a decelerating vehicle, C= centre of gravity

The direction of the net force should hit the track between front and rear axle; otherwise the vehicle will topple. The distance of front and rear axles from this direction of the net force are the leverages, which split the net force between front and rear axles inversely in proportion. In the figure, these leverages are marked ß and 1-ß. The larger share of the net force is exerted on the front wheels. Therefore, in classical cars without electronically controlled braking, usually brakes were adjusted so that about 75% of the power for braking is exerted on the front wheels. In modern cars, the pressures for braking at the wheels are controlled in such a way that the entire actual static frictional force at every single wheel is used to a maximum in an emergency brake. In figure 4, the fraction $\beta$ of the net force is split up to fractions $\beta$ of both the inertial force and gravitational force acting horizontally and vertically respectively. Likewise, at the rear wheels the $1-\beta$ of the net force is split up to fractions $1-\beta$ of both the inertial force and gravitational force acting horizontally and vertically respectively.

The fractions of gravitational force are in equilibrium with the corresponding fractions of the normal force, and the fractions of the inertial force are in equilibrium with the corresponding fractions of the static frictional forces. Adding all horizontal components of the forces at front and rear axle will let $\beta$ and $1-\beta$ sum to 1 and we arrive at the same equation (1) considering that the static frictional forces cannot exceed a maximum. The same maximum value holds for acceleration as for deceleration.

For good commercial tires and a dry and clean track the coefficient of static friction is 1 and maximum deceleration then is the constant of gravity g. For very high speed, this limit will be increased because drag force adds to forces from the brakes.

The limits of equation (1) and (2) hold only for optimum split of the braking forces between front and rear axles, which is usually attained in modern cars by electronic controlling of the brakes.

With the maximum of possible deceleration in motorcycles or bicycles, the driver can shift the centre of gravity on a position of 45° above the point of contact between the front tire and the track, then the rear wheel becomes free of any part of the net force, neither inertial or gravity part of the forces will act on the rear wheel. Thus the driver can easily lift the rear wheel, a performance called a `Stopy` by motorcycle enthusiasts, using equilibrium of forces and an application of free body diagram (figure 5).

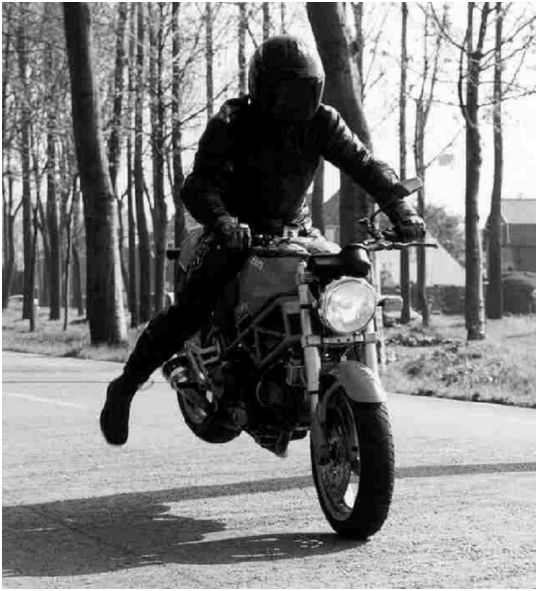

*Figure 5.     Braking of a motorcycle with maximum deceleration.*

If braking forces and inertial forces become more than the maximum of static frictional force, friction will fail and the tires will slide, locking the wheels and keeping them from rolling. Then the kinetic frictional force rather than static will be in equilibrium with inertial force. Since the kinetic frictional force coefficient is about 15% less than the static, deceleration will be diminished. Since the wheels are locked and kept from rolling, the car will slide in a straight fashion eventually out of a turn. In such an emergency brake the kinetic energy of the car will be transferred to heat not only in the braking system, but also in the tires causing screeching noise and considerable wear leaving skid marks on the road.

Thus for practical traffic maximum deceleration equals static friction coefficient µ´ times gravity g. Since µ has a practical maximum of 1, it is assumed that maximum deceleration equals gravity g, stronger braking would lock the wheels [Smith 2006, Unruh 1984, Risch 2002]

## 2      Limits to maximum speed of a vehicle

The free body diagram figure 1 also yields to an absolute limit of speed for vehicles driven by wheels. If a drag force of a vehicle cruising at constant speed replaces the inertial force from acceleration, the same balance of forces will hold. However, the drag force acts at the effective centre of the cross-section area of the vehicle taken perpendicular to the velocity while the inertial force acts at the centre of gravity S. To simplify the situation we assume both points to be at S. At very high speeds the airflow causes additional forces like lift force or negative lift, these are neglected here.

The drag force $F_D$ is:

$$F_D = A \cdot c \cdot \rho \cdot v^2 / 2$$

    A      cross-section area taken perpendicular to the velocity

    c      coefficient of drag

    $\rho$      density of the air

    v      velocity of the vehicle

The power needed to be supplied by the engine is:

$$P_D = v \cdot F_D$$

$$P_D = A \cdot c \cdot \rho \cdot v^3 / 2$$

Like with inertial force, the drag force is transferred to the wheels on the track and it is in equilibrium with static frictional force. The static frictional force, however, cannot exceed a maximum value and so does the drag force

$$F_{D\,max} = A \cdot c \cdot \rho \cdot v^2_{max} / 2 = F'_{max} = \mu' \cdot m \cdot g$$

A maximum value of speed can be derived

$$v^2_{max} = 2 \cdot \mu' \cdot m \cdot g / (A\,c\,\rho)$$

For example:

| | |
|---|---|
| cross-section area taken perpendicular to the velocity | A = 2 m² |
| drag coefficient | c = 0,5 |
| density of the air | $\rho$ = 1 kg/m³ |
| mass of the vehicle | m = 1000 kg |
| coefficient of static friction | µ´ = 1 |

maximum value of speed

$$v^2_{max} = 2 \cdot 1000 \text{ kg} \cdot 9{,}81 \text{m/s}^2 / (2\text{m}^2 \cdot 0{,}5 \cdot 1 \text{ kg/m}^3) = 19620 \text{ m}^2/\text{s}^2$$

$$v_{max} = 140 \text{ m/s} = 504 \text{ km/h}$$

At this maximum value of speed the front wheels become free of force, the vehicle thus becomes unstable and no steering is possible. Therefore, to attain such high speeds avoiding instability the drag coefficient has to be diminished considerably, and - or rocket drive has to be used rather than wheel drive. Constructing a car it is usual to shape the body so that negative lift forces from airflow are such that the car can be stable and steered well even at high speed. Following from these reasons the speed of limousines with high power has to be limited to 250km/h electronically.

## 3    Limits to maximum speed of a vehicle in a turn

The free body diagram of figure 1 can also be used to derive a maximum speed of a vehicle in a turn. The inertial force of figure 1 has to be replaced by the centripetal force, and the plane of drawing has to be turned by 90° such that the front view of the vehicle is shown in figure 6. Then a corresponding equilibrium of forces will hold.

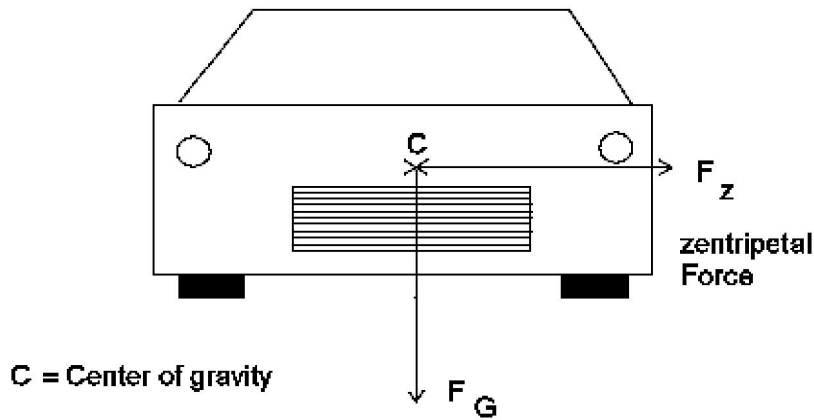

*Figure 6. Forces on a vehicle taking a turn in front view, gravitational force $F_G$ and centripetal force $F_Z$ at centre of gravity C form a net force transferred predominantly to the outer wheels.*

Like inertial force (in figure 1), the centripetal force is transferred to the outer wheels on the track (in figure 6) and it is in equilibrium with static frictional force. The static frictional force, however, cannot exceed a maximum value as for the centripetal force.

$$F'_{max} = \mu' \cdot F_G = \mu' \cdot m \cdot g = m \cdot v_{max}^2/r$$

Since the mass m cancels out, we arrive at:

$$v_{max}^2 / r = \mu' \cdot g$$

Thus the maximum speed in a turn is determined by the coefficient of static friction $\mu'$. For very good conditions of tires and track, $\mu'$ can become 1 and the maximum speed - calculated by:

$$v_{max}^2 = g \cdot r$$

This requires a suited position of the centre of gravity C that shall be not higher in elevation than half of the track width. Otherwise the vehicle will topple taking a turn.

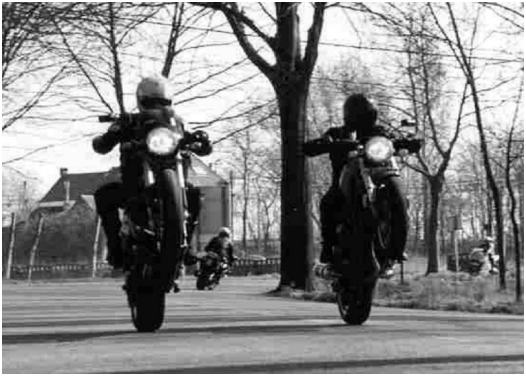

*Figure 7. Full acceleration and a fast turn on motorcycles, photo courtesy of a Dutch motorcycle club.*

## 4      Limits to acceleration and deceleration of a vehicle in a turn

When both inertial forces from acceleration or deceleration and centripetal forces are in action, both forces have to be added to a net force, which is transferred to the wheels and act on the track there. This net force is in equilibrium with the static frictional force that, however, cannot exceed a maximum value as with the net force.

The addition of the vectors of inertial forces from acceleration or deceleration and centripetal forces can be drawn in a coordinate system of accelerations. The maximum net acceleration is the same in every direction in this coordinate system and can be seen as a circle in figure 8.

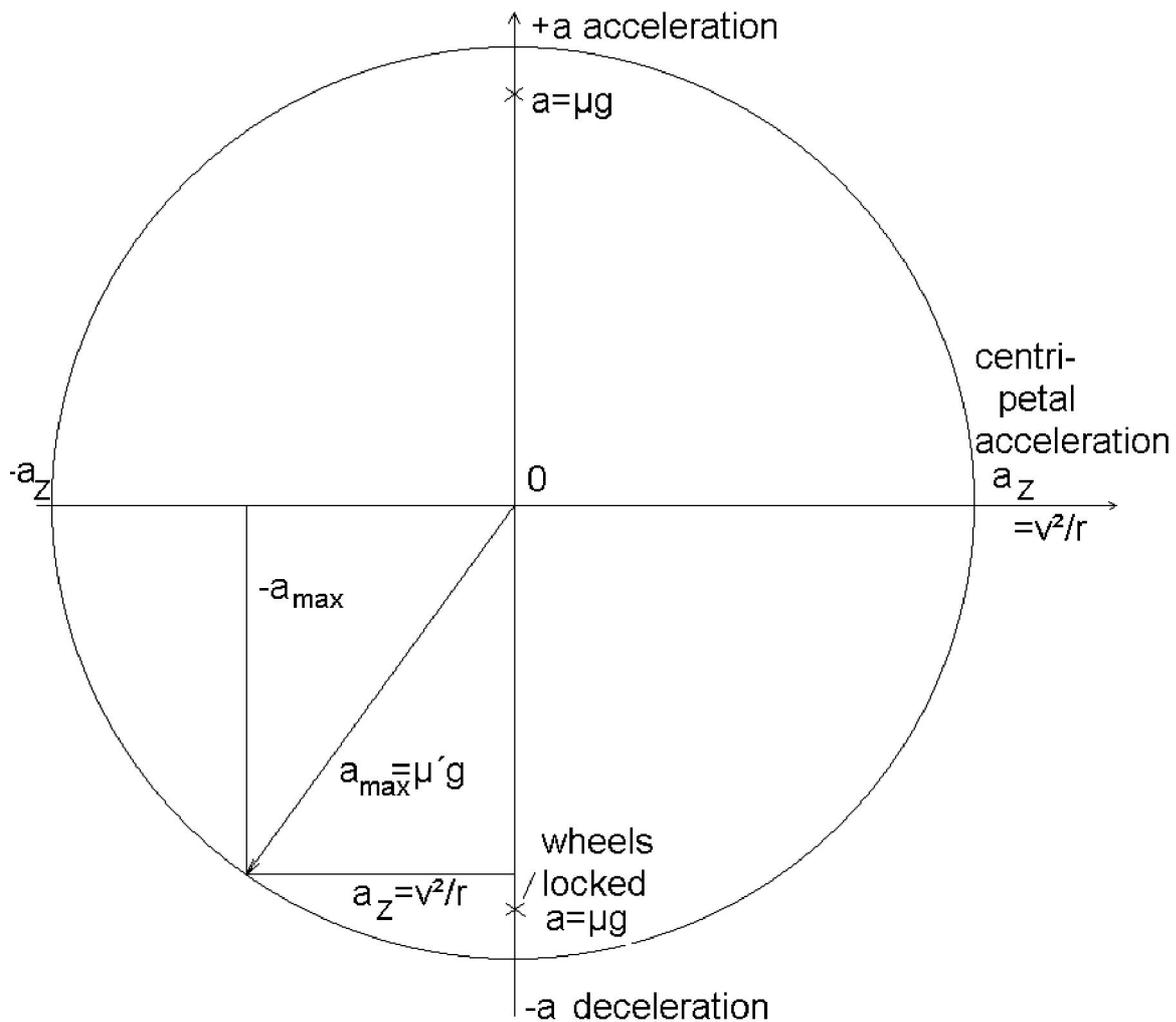

*Figure 8.   Kamm´s Circle showing the plane of accelerations for wheel driven vehicles.*

This circle is usually called Kamm´s circle, named after the head designer of an automobile factory north of the Alps that was noted for superior race cars. This circle describes acceleration vectors or likewise force vectors if multiplied by m. It is important to tell the students it does not display velocities, constant velocity is situated at the origin (zero). Horizontal direction displays centripetal acceleration, vertical direction upwards displays tangential acceleration and vertical direction downwards displays tangential deceleration. Accelerations are summed as vectors to net acceleration. In order to acquire force equilibrium, the amount of net acceleration has to be less or equal to gravity multiplied with coefficient of static friction $\mu´\cdot g$. That product $\mu´\cdot g$ is displayed as the circle. Possible conditions of driving are inside the circle, impossible conditions of driving outside the circle dividing these two realms. As soon as this circle boundary is trespassed, friction fails and the car will slide out of the turn in a tangential path.

When the car takes a turn, it is submitted to a centripetal acceleration. When braking, it cannot have all the total acceleration less or equal to gravity multiplied with coefficient of static friction µ´g since the centripetal acceleration has consumed already part of the total (net) acceleration. When adding or subtracting accelerations students have to keep in mind that they are perpendicular to each other and squares add or subtract like in Pythagoras's law (advanced students can be told these are vectors)

$$a_{dec\ max}^2 = a_{net}^2 - a_c^2 \qquad (3)$$

$$(a_{dec\ max})^2 = (µ´ \cdot g)^2 - (v^2/r)^2 \qquad (4)$$

Accelerations close to the circle can be attained by electronically controlled brakes and drive (ABS and ASR) only. When the circle is trespassed, static friction is replaced by dynamic friction, with sliding along the road leaving skid marks and an acceleration or deceleration 15% less since the wheels are locked and kept from rolling in a tangential movement in the car. Crosses in the diagram mark these states of driving.

## 5    Examples for maximum deceleration in a turn

Students can work on an example in teams (called Harl, Yama, Suzu, etc.) in a contest. A motorcycle with v = 50 km/h = 14 m/s on a wet road with a coefficient of static friction µ´= 0,5 drives through a flat turn with r = 50 m. How much is the maximum deceleration avoiding failure of friction and keeping the vehicle from sliding?

Velocity of vehicle             v = 50 km/h = 14 m/s = 31,3 mi/ h

centripetal acceleration        $a_Z = v^2/r = (14\ m/s)^2 / 50\ m = 4\ m/s^2$

maximum deceleration $\quad (a_{d\ max})^2 = (\mu'g)^2 - a_c^2 = (5\ m/s^2)^2 - (4\ m/s^2)^2 = (3\ m/s^2)^2$

This maximum 3 m/s² holds for both deceleration and instantaneous acceleration and can be attained only with electronic controls (ABS, ASR).

## 6  Consequences from maximum deceleration in a turn on handling of a vehicle

The concept of failing friction making a vehicle slide has some consequences for everyday handling of a vehicle. In turns taken swiftly, because of driving, braking, and centripetal forces at some of the wheels, a vehicle can get near the circular boundary in the diagram above. As a result of driving, braking, and centripetal forces transmitted to the wheels, for short instances the outer wheels can reach the circular border and beyond because of road surface roughness causing intermittent sliding of that wheel, which will leave its circular path and enter a tangential path for a short time. The consequences of rear wheel drive are contrary to that of front wheel drive, as can be seen from figure 9.

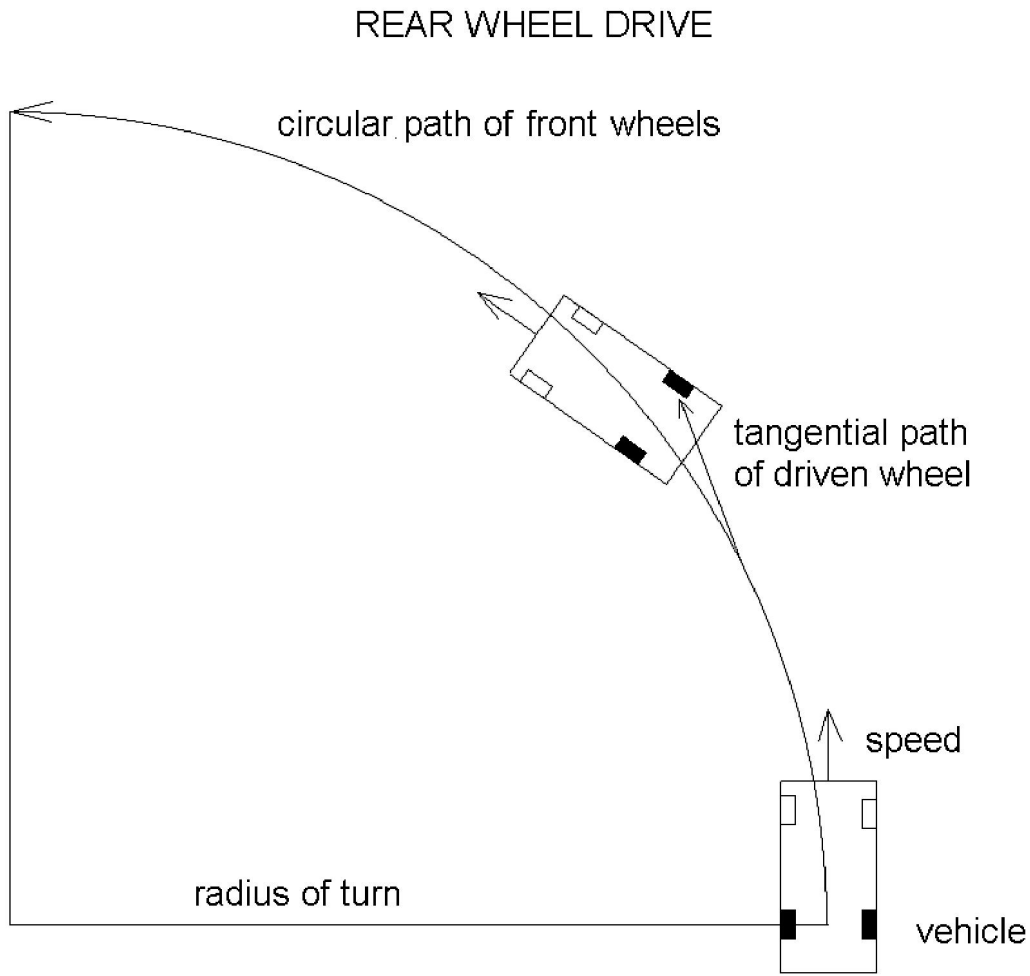

*Figure 9.    Handling of a vehicle with rear wheel drive.*

With rear wheel drive, the driven wheel on the outside of the turn has to transmit the driving force in addition to centripetal force, therefore it falls astray of its circular path towards a tangential path for short instances. Therefore, the whole vehicle turns slightly towards the inside of the turn. This type of handling is called „oversteering" by motorists.

The same concept and corresponding considerations can be made for front wheel drive, figure 10.

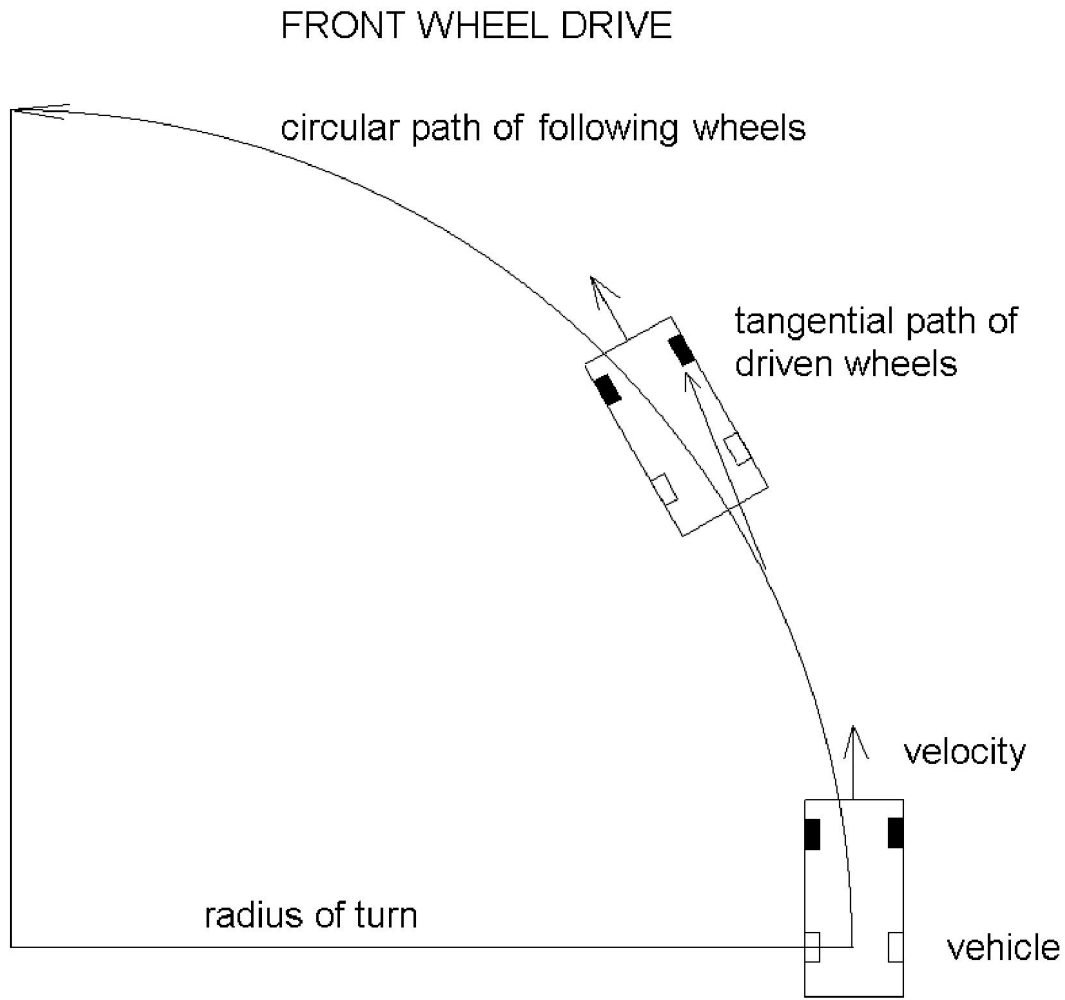

*Figure 10.    Handling of a vehicle with front wheel drive.*

With front wheel drive, the driven front wheel on the outside of the turn has to transmit the driving force in addition to centripetal force, therefore it falls astray of its circular path towards a tangential path for short instances. Therefore, the whole vehicle turns slightly towards the outside of the turn. This type of handling is called „understeering" by motorists.

# 7 Conclusions

Explanation of free body diagrams and equilibrium of forces applied to vehicle dynamics has been demonstrated to increase lecture attendance and participation in homework of 15 to 45 percent in undergraduate science and engineering classes. Engineering students in their first term had been divided in groups by random. For example, a home work about vibrations was done by 37% of the students in one group, in another group, however, the same homework was explained to be applied to a motorcycle seat in a curve and was done by 62% subsequently. The concepts of free body diagrams and equilibrium of forces are better understood by students working practical examples on vehicles of the associated theoretical problems. This is proved by the competency with which students solved associated problems. However, some of the students turned out to show difficulties to transfer the concepts of free body diagrams and equilibrium of forces gained with consideration of vehicles to different mechanics problems such as weights at ropes or cantilevers.

FIGURE CAPTIONS

*Figure 1. Simplified free body diagram of a vehicle with full acceleration, C= centre of gravity.*

*Figure 2. Starting of a car with maximum acceleration.*

*Figure 3. Full acceleration on a motorcycle (courtesy of a Belgian motorcycle club).*

*Figure 4. Free body diagram of a decelerating vehicle, C= centre of gravity*

*Figure 5. Braking of a motorcycle with maximum deceleration.*

*Figure 6. Forces on a vehicle taking a turn in front view, gravitational force $F_G$ and centripetal force $F_Z$ at centre of gravity C form a net force transferred predominantly to the outer wheels.*

*Figure 7. Full acceleration and a fast turn on motorcycles, photo courtesy of a Dutch motorcycle club*

*Figure 8. Kamm´s Circle showing the plane of accelerations for wheel driven vehicles.*

*Figure 9. Handling of a vehicle with rear wheel drive.*

*Figure 10. Handling of a vehicle with front wheel drive.*